\documentclass[aps,prb,twocolumn,footinbib,superscriptaddress]{revtex4-2}

\usepackage{amsmath}
\usepackage{amssymb}
\usepackage{siunitx}
\usepackage{graphicx}
\usepackage{xspace}
\usepackage{braket}
\usepackage{bm}
\usepackage{placeins}
\usepackage{leftidx}
\usepackage[usenames,dvipsnames]{color}
\usepackage[utf8]{inputenc}
\usepackage{natbib}
\usepackage[pdftex, colorlinks=true, urlcolor=blue, linkcolor=blue, citecolor=blue, pdfstartview=FitH, plainpages=false]{hyperref}
\usepackage{chemformula}
\usepackage{float}
\usepackage{todonotes}

\begin{document}

\preprint{APS/123-QED}
\title{From Symmetry to Stability: Structural and Electronic Transformation in \ch{Cs2KInI6}}
\date{\today}

\author{Mohammad Bakhsh}
\email{mohammad.bakhsh@uclouvain.be}
\affiliation{UCLouvain, Institute of Condensed Matter and Nanosciences (IMCN), Chemin des \'Etoiles~8, B-1348 Louvain-la-Neuve, Belgium}
\author{Victor Trinquet}
\affiliation{UCLouvain, Institute of Condensed Matter and Nanosciences (IMCN), Chemin des \'Etoiles~8, B-1348 Louvain-la-Neuve, Belgium}
\author{Rogério Almeida Gouvêa}
\affiliation{UCLouvain, Institute of Condensed Matter and Nanosciences (IMCN), Chemin des \'Etoiles~8, B-1348 Louvain-la-Neuve, Belgium}
\author{Gian-Marco Rignanese}
\affiliation{UCLouvain, Institute of Condensed Matter and Nanosciences (IMCN), Chemin des \'Etoiles~8, B-1348 Louvain-la-Neuve, Belgium}
\affiliation{WEL Research Institute, avenue Pasteur 6, 1300 Wavre, Belgium.}
\author{Samuel Ponc\'e}
\email{samuel.ponce@uclouvain.be}
\affiliation{UCLouvain, Institute of Condensed Matter and Nanosciences (IMCN), Chemin des \'Etoiles~8, B-1348 Louvain-la-Neuve, Belgium}
\affiliation{WEL Research Institute, avenue Pasteur 6, 1300 Wavre, Belgium.}


\begin{abstract}
Cs$_2$KInI$_6$ is a promising lead-free halide double perovskite with a calculated direct band gap of 1.94~eV, ideal for solar cell applications. 
Our first-principles calculations reveal that its cubic phase (Fm$\bar{3}$m) is dynamically unstable. 
Using an accelerated machine learning approach, we identify 42 dynamically stable structures and further validate these findings using first-principles calculations on 11 of these. 
The most stable phase has Cmc$2_1$ symmetry with 20 atoms/unit cell. 
It lies 13~meV/atom above the convex hull but lacks octahedral cation coordination. 
The most stable perovskite-like structure has P$\bar{3}$ symmetry with 10 atoms/unit cell and low octahedral connectivity. 
Structure-property trade-offs are highlighted, with calculated distortions generally widening the band gap, shifting it from direct to indirect, and flattening the band edges. 
This work showcases the synergy of genetic algorithms, machine-learned potentials, and first-principles validation for discovering stable, complex materials.
\end{abstract}

\maketitle

Halide perovskites have attracted significant attention for solar cell applications due to their tunable band gap~\cite{miah2024band}, high absorption coefficient~\cite{de2014organometallic} and low-cost solution-based processability~\cite{chilvery2016perspective}. 
As a result, the power conversion efficiency of single-junction halide perovskite solar cells has increased rapidly, reaching 27\% in a short period~\cite{nrel2025efficiency}. 
However, most high-performance halide perovskites, such as the organic–inorganic hybrid methylammonium (MA) lead halides~\cite{zhao2016organic,Ponce2019,Xia2021} and inorganic cesium lead halides~\cite{eperon2015inorganic}, contain toxic lead, which poses serious environmental and health concerns that hinder large-scale commercialization~\cite{lyu2017addressing}. 
In addition, these materials often suffer from limited environmental stability~\cite{kye2018critical,jong2018influence}. 
These challenges have prompted researchers to explore the broader chemical space in search of lead-free and more stable halide perovskite alternatives.
In particular, lead-free halide double perovskites with the general formula A$_2$MM$^\prime$X$_6$ have emerged as promising candidates for photovoltaic applications~\cite{slavney2016bismuth, volonakis2016lead, savory2016can}. 
These structures are derived by doubling the formula unit of the conventional single perovskite APbX$_3$ (X being an halogen) and replacing the two lead cations with two different metal cations, M and M$^\prime$.
These substituting cations may both adopt a +2 oxidation state, or one may be +1 and the other +3, so that their combined charge is equivalent to that of two Pb$^{2+}$ ions.
Ideally with a $ns^2$-type cation, which typically shows enhanced optical absorption, well-balanced carrier transport, good defect tolerance, and long carrier diffusion lengths~\cite{Li2021}.
This substitution scheme significantly expands the accessible chemical space compared to APbX$_3$ perovskites.
Such perovskites have been reported to exhibit improved thermal and moisture stability compared to hybrid perovskites such as (MA)PbI$_3$.
In particular, Cs$_2$AgBiBr$_6$ demonstrates improved durability and defect tolerance~\cite{slavney2016bismuth}. 

However, it has an indirect band gap~\cite{mcclure2016cs2agbix6}, which limits its effectiveness in photovoltaic applications~\cite{savory2016can}.
To find better candidates, Cai~\emph{et~al.}~\cite{cai2019high} performed a high-throughput study and reported the Cs$_2$KInI$_6$ perovskite but filtered it out due to its high energy above the convex hull (56~meV/atom)~\cite{jain2013commentary}. 
In a more recent work, its cubic phase was identified as a promising candidate for photovoltaic applications among 1026~compounds~\cite{qi2023optoelectronic} due to its direct band gap that matches optimal sunlight absorption.
Nevertheless, its vibrational dynamical stability was not investigated.
In this work, we study the dynamical stability of \ch{Cs2KInI6} in the cubic phase with space group Fm$\bar{3}$m (225). 
Our phonon analysis reveals that this high-symmetry phase is dynamically unstable, leading to 42 more stable phases with lower symmetry.  
In the cubic phase, both K and In cations occupy octahedral coordination environments. 
These octahedra are corner-sharing, forming a three-dimensional alternating network. 
The dynamically stable phases are analyzed on the basis of the possible modification of the local environments of the K and In cations and the polyhedra connectivity, which has a significant impact on transport properties~\cite{Hoye2021}.
Furthermore, the impact of these structural transformations on electronic properties is evaluated to estimate the suitability of this material for photovoltaic applications.

The initial crystal structure of cubic \ch{Cs2KInI6} was obtained from the Materials Project database~\cite{jain2013commentary}. 
Electronic structure calculations were performed using density functional theory (DFT)~\cite{hohenberg1964inhomogeneous,kohn1965self} as implemented in the \textsc{Quantum ESPRESSO} (QE) package~\cite{Giannozzi2017}. 
The Perdew–Burke–Ernzerhof (PBE)~\cite{perdew1996generalized} exchange-correlation functional was used with norm-conserving pseudopotentials from the \textsc{PseudoDojo} library~\cite{van2018pseudodojo}.
For three structures (Fm$\bar{3}$m, P$\bar{3}$ and I$\bar{4}$2m), each containing 10 atoms per unit cell, the band structures were also computed using the hybrid Heyd--Scuseria--Ernzerhof (HSE06)~\cite{Heyd2003} functional on a homogeneous grid and Wannier interpolated~\cite{Pizzi2020} to produce the band structure.
After performing convergence tests on total energy per atom, lattice constants, and atomic positions, the plane-wave kinetic energy cutoff was set to 80~Ry. 
The convergence thresholds were chosen to be 0.1~mHa/atom for total energy and 0.1\% of the converged value for lattice constants and atomic positions.
The Brillouin zone was sampled using a 3$\times$3$\times$3 $\Gamma$-centered Monkhorst–Pack~\cite{monkhorst1976special} \textbf{k}-point mesh. 
Structural optimization was performed with a force convergence threshold of $10^{-5}$~Ry/bohr. 
Phonon calculations were performed using density functional perturbation theory (DFPT)~\cite{Gonze1997,baroni2001phonons} for high-accuracy validation, using \textbf{q}-point grids with a density below (0.5/\text{\AA})$^{3}$~\cite{Bercx2025} and a \textbf{k}-point mesh denser or equal to the corresponding \textbf{q}-point grid. 
For high-throughput analysis, phonons were computed using the finite-displacement method with a machine-learned interatomic potential (MLIP) based on the message-passing atomic cluster expansion (MACE), specifically the MACE-OMAT-0 model~\cite{Batatia2022mace,Batatia2025,Loew2025}, with phonon frequencies converged to within 2~cm$^{-1}$.
The energies above the convex hull of the \ch{Cs2KInI6} phases were calculated with respect to their stable decomposition products identified from the Materials Project database, using the \textsc{pymatgen}~\cite{Ong2013} phase diagram module together with total energies recomputed with \textsc{Quantum ESPRESSO}. 
Structural analysis was performed using \textsc{ChemEnv}~\cite{waroquiers2020chemenv} to determine the coordination environments of K and In cations.

\begin{figure}[t] 
    \centering
    \includegraphics[width=0.95\linewidth]{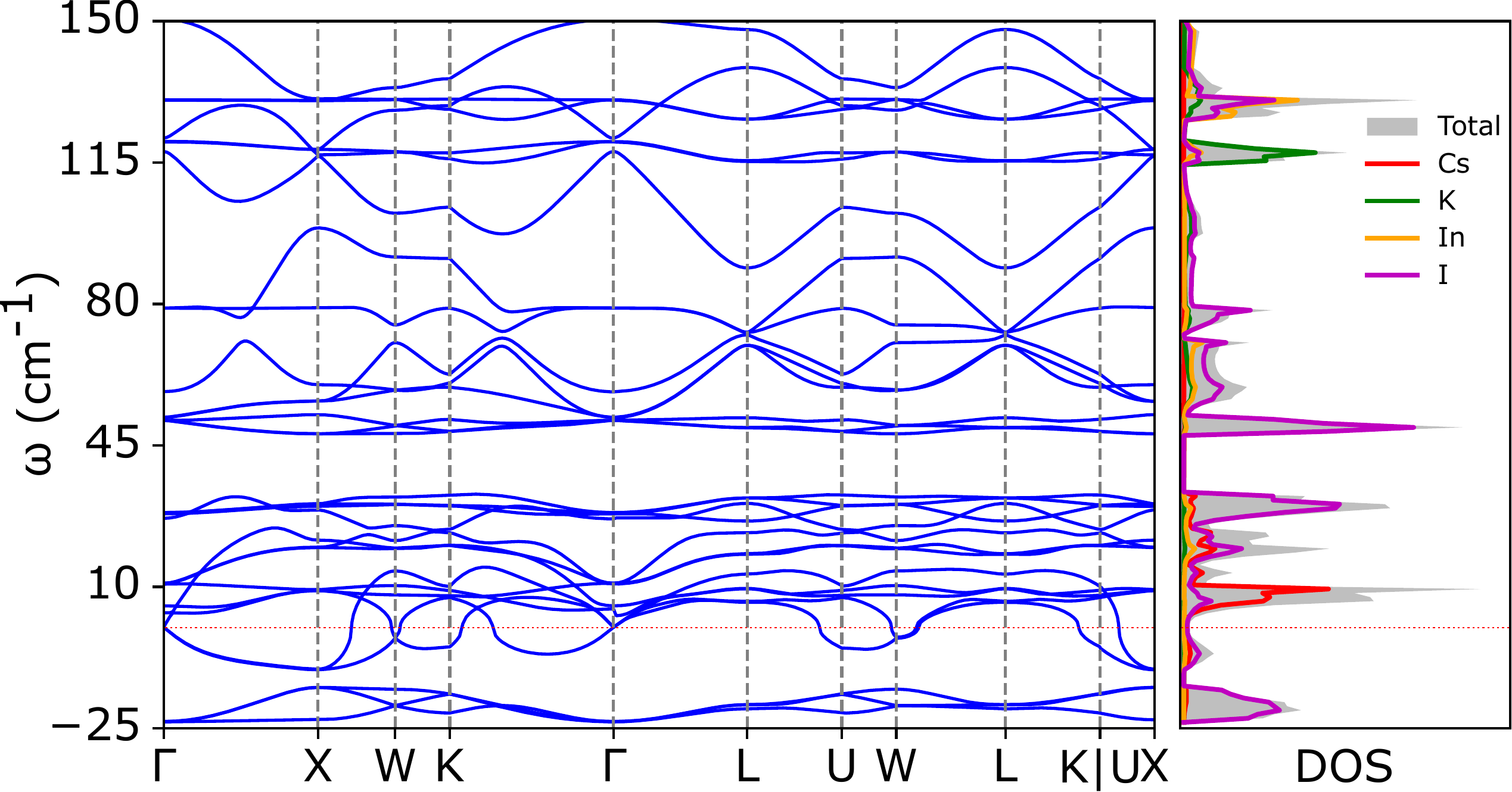}
    \caption{\label{fig:1}  
     Phonon dispersion and phonon density of states (DOS) for cubic \ch{Cs2KInI6} (space group Fm$\bar{3}$m, 225), computed using density functional perturbation theory with a $3 \times 3 \times 3$ \textbf{k}-point grid and $2 \times 2 \times 2$ \textbf{q}-point grid. 
    }
\end{figure}

Figure~\ref{fig:1} shows the phonon dispersion and phonon density of states (DOS) of cubic \ch{Cs2KInI6} using DFPT. 
The presence of unstable modes indicates that this high-symmetry phase lies at a saddle point on the potential energy surface~\cite{jong2019anharmonic}, reflecting its dynamical instability.
Although such an instability can sometimes be stabilized by including anharmonic effects at finite temperature~\cite{Yin2025}, it is useful to identify the ground-state structure from DFT to describe perovskites at cryogenic temperatures or in situations where the high-symmetry phase cannot be stabilized.
To achieve this, a MACE potential was used with our in-house code called \textsc{VibroML}~\cite{Rogerio2026vibroml} to explore the complex potential energy surface of the \ch{Cs2KInI6} structure. 
To resolve its dynamical instabilities, we use a genetic algorithm for a global search aimed at identifying dynamically stable low-energy polymorphs in the energy landscape of the material. 
The procedure begins by identifying the soft phonon modes of the unstable cubic phase. 
An initial population of distorted candidate structures is then generated by displacing atoms along the eigenvectors of these unstable modes. 
This initial generation explores various displacement amplitudes and coupling with other soft modes to ensure a diverse search space.
The algorithm then proceeds through an iterative evolutionary cycle. 
At each generation, the fitness of every candidate structure is evaluated on the basis of its total energy per atom after a structural relaxation using the MACE potential.
The candidates with the lowest energies are preferentially selected as parents for the next generation. 
New offspring structures are created using genetic operators such as crossover, which combines features from parent structures, and mutation, which applies small, random perturbations. 
This process, inspired by natural selection, effectively navigates the energy landscape toward more stable configurations.

\begin{table*}[!htbp]
    \caption{\label{tab:1}
    Lattice parameters ($a$, $b$, and $c$ in \AA; angles $\alpha$, $\beta$, and $\gamma$ in $^\circ$) of the conventional unit cell, number of atoms in the primitive unit cell ($N_\mathrm{at}$), and energy above the convex hull ($E_{hull}$) for the dynamically stable DFPT polymorphs of \ch{Cs2KInI6} in meV/atom.}
    \begin{ruledtabular}
    \begin{tabular}{l r c r r r r r r c r}
         \multicolumn{2}{c}{Space group} & id & \multicolumn{1}{c}{$a$} & \multicolumn{1}{c}{$b$} & \multicolumn{1}{c}{$c$} & \multicolumn{1}{c}{$\alpha$} & \multicolumn{1}{c}{$\beta$} & \multicolumn{1}{c}{$\gamma$} & $N_\mathrm{at}$ & $E_{hull}$ \\
    \hline
    \rule{0pt}{3ex}Fm$\bar{3}$m & 225 &00 & 12.669 & 12.669 & 12.669 & 90.0  & 90.0  & 90.0 & 10 & \ 55  \\
    P$\bar{3}$ & 147 &05 & 8.649 & 8.649 & 8.649 & 90.0 & 90.0 & 120.0 & 10 & 24 \\
    I$\bar{4}$2m & 121 &07 & 7.756 & 7.756 & 19.018 & 90.0 & 90.0 & 90.0 & 10 & 21  \\
    Cmc$2_1$ & 36 &37& 10.644 & 19.482 & 10.299 & 90.0 & 90.0 & 90.0 &20 & 13  \\
    P$\bar{1}$ & 2 &42 & 17.427 & 17.502 & 17.662 & 91.7 & 119.0 & 119.1 & 80 & 25 \\       
    \end{tabular}
    \end{ruledtabular}
\end{table*}
\begin{figure*}[!htbp]
    \centering
    \includegraphics[width=0.85\linewidth]{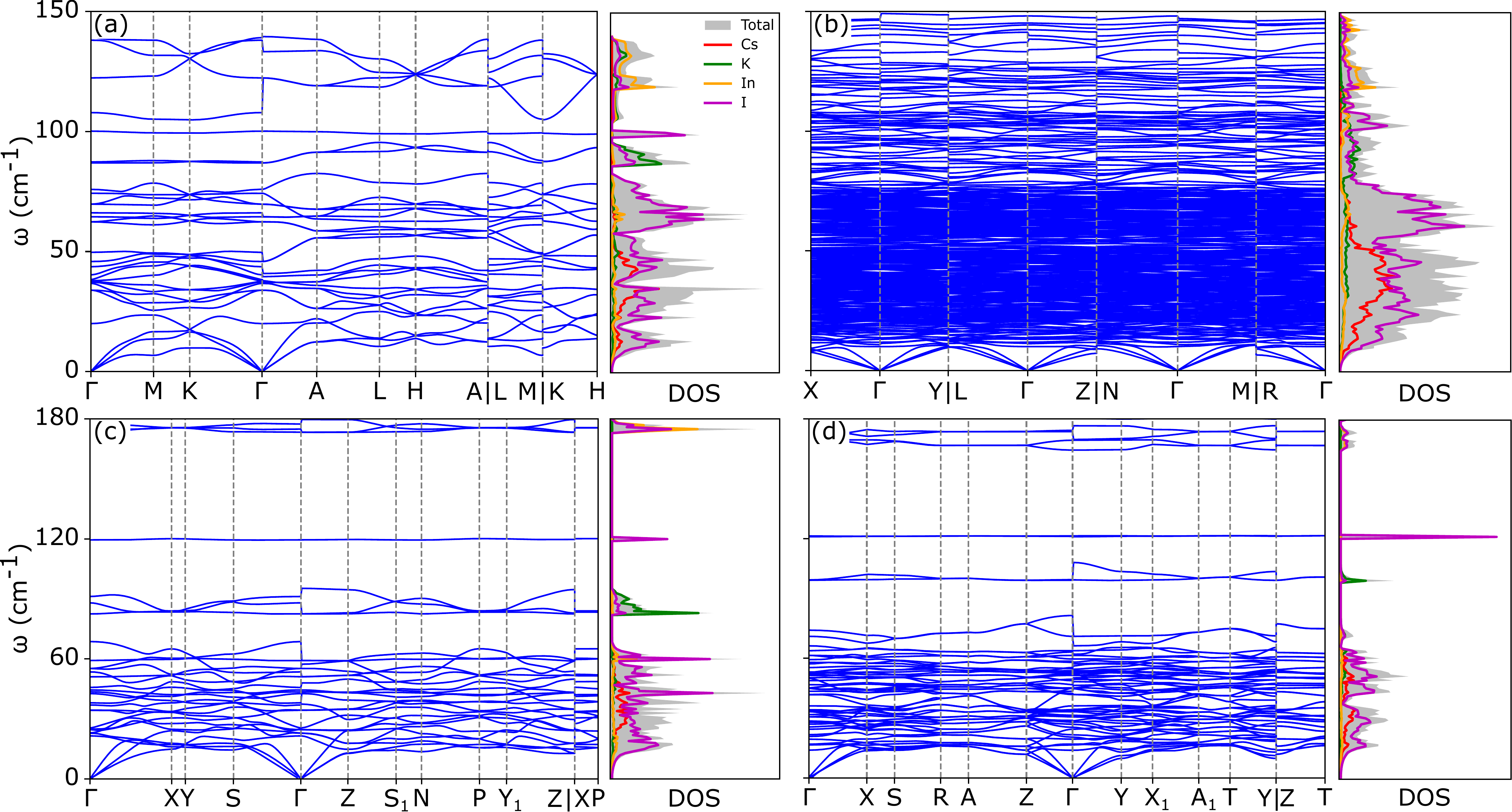}
     \caption{\label{fig:2}
     DFPT phonon calculations with density of states (DOS) of four \ch{Cs2KInI6} phases. 
     (a) P$\bar{3}$ with $4 \times 4 \times 4$ \textbf{k}-grid and $2 \times 2 \times 2$ \textbf{q}-grid. 
     (b) P$\bar{1}$ with $2 \times 2 \times 2$ \textbf{k}-grid and $1 \times 1 \times 1$ \textbf{q}-grid.
     (c) I$\bar{4}$2m with $3 \times 3 \times 3$ \textbf{k}-grid $2 \times 2 \times 2$ \textbf{q}-grid. 
     (d) Cmc$2_1$ with $3 \times 3 \times 3$ \textbf{k}-grid and $2 \times 2 \times 2$ \textbf{q}-grid.}
\end{figure*}

Crucially, the workflow incorporates a feedback loop: a phonon calculation is performed on the most promising low-energy candidates of each generation. 
If any instabilities persist, the corresponding soft modes are used to guide the creation of new distorted structures in subsequent iterations. 
This evolutionary approach allows for a broad and efficient search that systematically uncovers lower-energy, dynamically stable polymorphs originating from the unstable parent structure.
This process yielded 42 new polymorphs of \ch{Cs2KInI6}, which were found to be dynamically stable based on frozen-phonon calculations using MACE (see Table~S1 of the Supplemental Information, SI~\cite{SupplementalMaterial}).
They can be classified according to the number of atoms in their primitive unit cell (10, 20, 40, and 80 atoms).
However, confirmation using first-principles DFPT calculations is required to validate the dynamical stability predicted by MACE.
Therefore, we selected all polymorphs with 10 atoms, two with 20 atoms (one perovskite structure and the most stable non-perovskite structure), one with 40 atoms, and the one for the 80 atoms count.
This resulted in a total of 11 structures for which DFPT phonon calculations were performed.
We found that in 9 cases the DFPT calculations confirmed the dynamic stability predicted by MACE, whereas in two cases MACE predicted stability but DFPT revealed dynamical instabilities. As a result, DFPT validation remains necessary for the remaining 31 structures. However, these structures contain 20 and 40 atoms in their primitive unit cells and therefore require substantial computational resources for DFPT phonon calculations, making a complete DFPT validation of all candidate polymorphs computationally demanding. 
In addition, some polymorphs of \ch{Cs2KInI6} were predicted to be dynamically unstable by both MACE and DFPT; these structures are listed in Table~S2~\cite{SupplementalMaterial}.
Overall, the DFPT phonon dispersions and phonon densities of states show very good agreement with the corresponding MACE results (see Figs.~S1–S3 and S5–S6 for phonon dispersions, and Figs.~S4 and S7 for phonon densities of states in the SI~\cite{SupplementalMaterial}).
For the remainder of this work, we focus on four structures (IDs = 05, 07, 37, and 42), which are confirmed to be dynamically stable at both the MACE and DFPT levels. 
The P$\bar{3}$ (147) structure (id=05) with 10~atoms per unit cell and a double perovskites structure; 
the I$\bar{4}$2m (121) structure (id=07) which is the most stable structure with 10~atoms and high symmetry;
the Cmc$2_1$ (36) structure (id=37) as the most stable; and 
the P$\bar{1}$ (2) structure (id=42) which is the largest structure with 80~atoms per unit cell.
We report their lattice parameters, number of atoms in the primitive unit cell, and the energy above the convex hull ($E_{hull}$) in Table~\ref{tab:1}.
The lattice parameters were recomputed using PBEsol exchange-correlation functional, which are reported in Table~S3 of SI~\cite{SupplementalMaterial}. The lattice vectors obtained with PBEsol are on average 3.4\% shorter than those obtained with PBE. 
According to their energies above the convex hull, the P$\bar{3}$ (147), I$\bar{4}$2m (121), Cmc$2_1$ (36), and P$\bar{1}$ (2) phases are thermodynamically metastable. However, their $E_{hull}$ values lie within 25~meV/atom, which is the typical threshold for the experimentally observed iodides~\cite{Sun2016}.   
We also present their X-ray diffraction in Fig.~S8 of the SI~\cite{SupplementalMaterial} which show large differences that can easily be resolved experimentally.
We also present their phonon dispersion and phonon DOS in Fig.~\ref{fig:2}.
Remarkably, the two perovskites structures, Figs.~\ref{fig:2}(a) and (b), show a continuous spectra with phonon frequencies below 150~cm$^{-1}$ while the two non-perovskite structures show multiple localized and high-energy vibrational modes up to 180~cm$^{-1}$, see Figs.~\ref{fig:2}(c) and (d). 
In particular, both show the same iodine localized vibrational peak at 120~cm$^{-1}$ which is characteristic of these structures.
These differences can be explained by studying their crystal structure, shown in Fig.~\ref{fig:3}.
The two structures shown in Figs.~\ref{fig:3}(c) and (d) cannot be classified as double perovskites as both cation sites (K and In in this case) are not in an octahedral coordination environments as the cubic phase of \ch{Cs2KInI6}. 
Indeed, the In cation is in a tetrahedral environment, while the geometry of the K cation is not well-defined; it is coordinated with six I atoms in I$\bar{4}$2m and seven I atoms in Cmc$2_1$. 
For the P$\bar{3}$ structure, our \textsc{ChemEnv} analysis gives 100\% octahedral similarity with a slightly distorted octahedron for the K cations. 
Finally, in the P$\bar{1}$ structure, the K cations do not exhibit an octahedral environment. 
However, \textsc{ChemEnv} reports a combination of different coordinations for each K site, see Fig.~S9 of the SI~\cite{SupplementalMaterial}.
From this analysis, the P$\bar{3}$ (147) phase is classified as a double perovskite while the case of the P$\bar{1}$ structure is more complex since the In cations are octahedrally coordinated while the K cations are not.
Since the P$\bar{3}$ structure is classified as a double perovskite, we compare it with the cubic phase. 
In the latter, the In and K octahedra share corners in all three directions, forming a three-dimensional alternating network, see Fig.~\ref{fig:3}(a). 
In the trigonal phase, they share faces instead of corners, see Fig.~\ref{fig:3}(b), producing one-dimensional strips in which the In and K octahedra alternate.
This change from a three-dimensional corner-sharing network to one-dimensional face-sharing strips reflects a significant reduction in structural connectivity.

\begin{figure*}[t] 
    \centering
    \includegraphics[width=0.7\linewidth]{Fig.3.pdf}
    \caption{\label{fig:3} 
     Crystal structures of (a) Fm$\bar{3}$m, (b) P$\bar{3}$, (c) I$\bar{4}$2m, (d) Cmc$2_1$ and (e)  P$\bar{1}$, displaying the coordination environments of In and K cations and the way they are connected.}
\end{figure*}

\begin{figure*}[!htbp]
    \centering
    \includegraphics[width=0.8\linewidth]{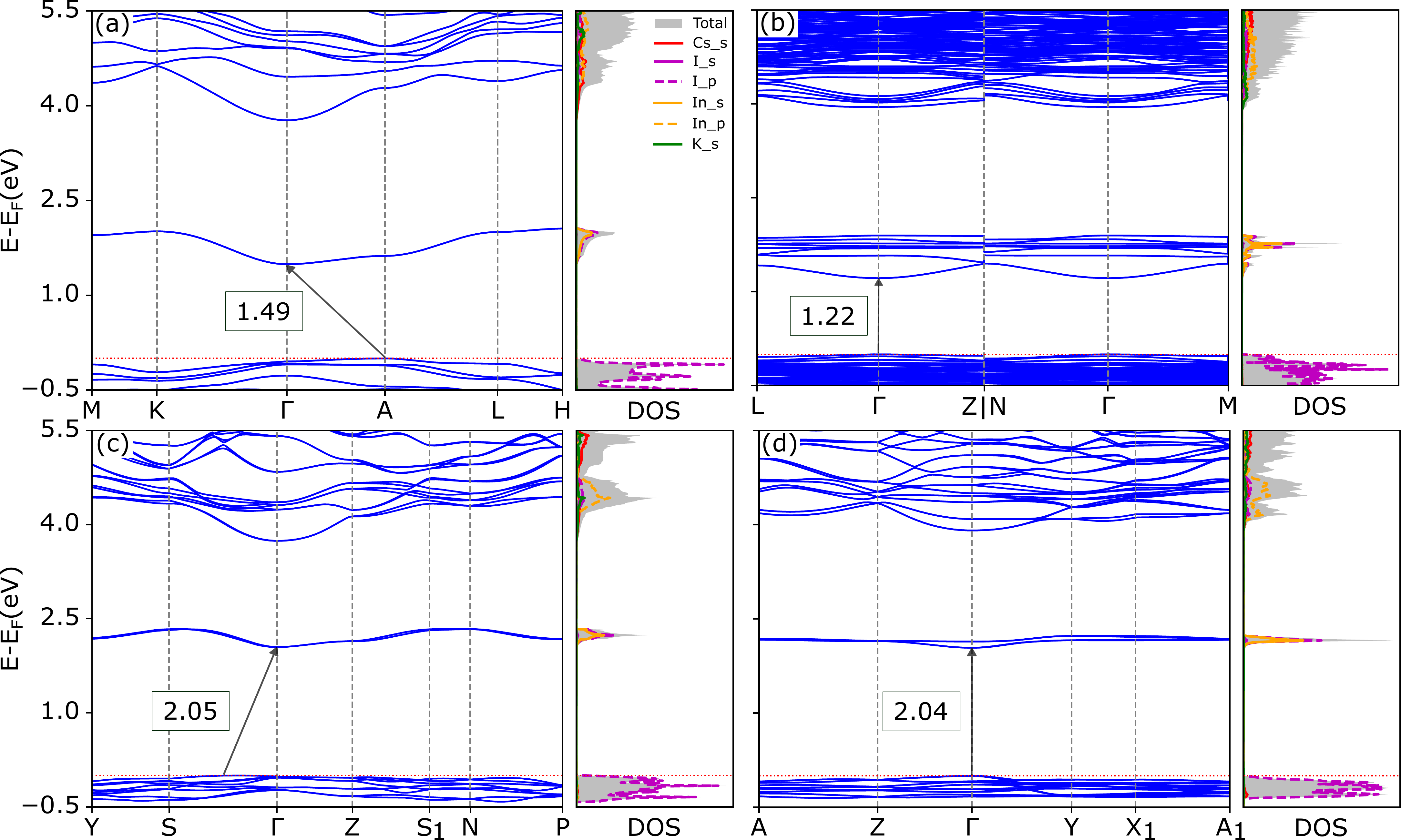}
    \caption{\label{fig:4} 
    Electronic band structures and densities of states of dynamically stable phases of \ch{Cs2KInI6} with PBE including spin-orbit coupling (SOC): (a) P$\bar{3}$, (b) P$\bar{1}$, (c) I$\bar{4}$2m, and (d) Cmc$2_1$.
    The energies are expressed with respect to the Fermi level ($E_{\rm F}$), located at the valence band maximum.} 
\end{figure*}

We now turn to the analysis of the impact of structural differences on electronic properties.
The cubic phase has a direct band gap at the $\Gamma$ point, with value of 1.24 eV at the PBE level (Fig.~S10(a)), reduced to 1.06 eV upon inclusion of SOC (Fig.~S10(b)), increased to 1.94 at the HSE06+SOC level, as shown in Fig.~S12(a) of the SI~\cite{SupplementalMaterial}. 
Due to structural distortions, most notably changes in the In and K coordination environment and the connectivity of the their octahedra, the band gaps of the four considered cases increase. 
This trend is evident from Fig.~\ref{fig:4}, which presents the PBE+SOC electronic band structures of the four investigated phases of Cs$_2$KInI$_6$.    
Moreover, the band gap changes from direct to indirect in two phases (P$\bar{3}$ and I$\bar{4}$2m).
The  corresponding PBE band structures without SOC are provided in Fig.~S11 of the SI~\cite{SupplementalMaterial}. 
The double perovskite structure, P$\bar{3}$ has an indirect band gap of 1.68 eV at the PBE level (Fig.~S11(a))~\cite{SupplementalMaterial}, 1.49 eV with PBE+SOC (Fig.~\ref{fig:4}(a)) and 2.29 eV with HSE06+SOC as presented in the Fig.~S12(b)~\cite{SupplementalMaterial}. 
This gap is only slightly larger than that of the cubic phase.
The structure P$\bar{1}$, in which only the In cations form octahedrons has a direct band gap of 1.35~eV at the PBE level (Fig.~S11(b))~\cite{SupplementalMaterial} and 1.22~eV with PBE+SOC, see Fig.~\ref{fig:4}(b). These band gap values are close to that of the cubic phase. Therefore, among the considered phases, P$\bar{1}$ could be a suitable candidate for the photovoltaic applications if synthesized.
The I$\bar{4}$2m and Cmc$2_1$ structures that are not classified as double perovskites have significantly larger band gaps than the cubic phase. I$\bar{4}$2m has band gap values of 2.24~eV at the PBE level (Fig.~S11(c))~\cite{SupplementalMaterial}, 2.05~eV with PBE+SOC (Fig.~\ref{fig:4}(c)), and 2.93~eV with HSE+SOC as reported in the Fig.~S12(c)~\cite{SupplementalMaterial}. The Cmc$2_1$ has a PBE band gap of 2.26~eV (Fig.~S11(d))~\cite{SupplementalMaterial} and 2.04~eV with PBE+SOC, see Fig.~\ref{fig:4}(d).
P$\bar{1}$ contains 80 atoms and Cmc$2_1$ contains 20 atoms. Therefore, their band structures could not be computed with HSE06+SOC level due to the high computational cost.
However, based on the HSE06+SOC band gaps of Fm$\bar{3}$m, P$\bar{3}$, and I$\bar{4}$2m phases, the band gap increases by approximately 60\% on average compared to the PBE+SOC values. Following this trend, the HSE06+SOC band gap of P$\bar{1}$ should be approximately 1.95~eV, while that of Cmc$2_1$ approximately 3.26~eV.
Overall, structural distortions increase the band gap and flatten the conduction and valence bands, thereby increasing the effective electron and hole masses, as reported in Table~\ref{tab:2}.

\begin{table}[t]
   \caption{\label{tab:2}
   Effective masses of electron and hole for the dynamically stable phases of Cs$_2$KInI$_6$ along with the high symmetry cubic phase. 
   The valence band maximum (VBM) and the conduction band minimum (CBM) are also reported. 
   For the I$\bar{4}$2m structure, the VBM does not lie at a high-symmetry point and is indicated with a star ($*$).}
    \begin{ruledtabular}
    \begin{tabular}{l r c c c c c}
        \multicolumn{2}{c}{Space group} & VBM & CBM & Direction & \textbf{$m_e^*$} & \textbf{$m_h^*$} \\
        \hline
        \rule{0pt}{3ex}Fm$\bar{3}$m
          & 225 & $\Gamma$ & $\Gamma$ & \makebox[1em]{$\Gamma$}--\makebox[1em]{L}                & 0.349 &  1.375 \\
          &     &          &          & \makebox[1em]{$\Gamma$}--\makebox[1em]{K}                & 0.349 &  1.648 \\
        P$\bar{3}$
          & 147 &  A       & $\Gamma$ & \makebox[1em]{$\Gamma$}--\makebox[1em]{A}                & 1.402 & ---  \\
          &     &          &          & \makebox[1em]{$\Gamma$}--\makebox[1em]{K}                & 0.548 & ---   \\
          &     &          &          & \makebox[1em]{A}--\makebox[1em]{$\Gamma$}                & ---   & 6.120  \\
          &     &          &          & \makebox[1em]{A}--\makebox[1em]{L}                       & ---   & 1.079  \\
        P$\bar{1}$
          &   2 & $\Gamma$ & $\Gamma$ & \makebox[1em]{$\Gamma$}--\makebox[1em]{Z}                & 0.517 & 2.418 \\
          &     &          &          & \makebox[1em]{$\Gamma$}--\makebox[1em]{L}                & 0.568 &  2.132 \\
        I$\bar{4}$2m
          & 121 &   $*$    & $\Gamma$ & \makebox[1em]{$\Gamma$}--\makebox[1em]{Z}                & 1.419 & --- \\
          &     &          &          & \makebox[1em]{$\Gamma$}--\makebox[1em]{S}                & 0.615& ---\\
          &     &          &          & \makebox[1em]{$*$}--\makebox[1em]{$\Gamma$}                & ---   & 7.804\\
          &     &          &          & \makebox[1.4em]{\phantom{$_1$}$*$}--\makebox[1.4em]{S}     & ---   &  7.922 \\
        Cmc2$_1$
          &  36 & $\Gamma$ & $\Gamma$ & \makebox[1em]{$\Gamma$}--\makebox[1em]{Y} & 0.724 & 0.607 \\
          &     &          &          & \makebox[1em]{$\Gamma$}--\makebox[1em]{Z}                & 1.111 & 2.233 \\
    \end{tabular}
    \end{ruledtabular}
\end{table}

We investigate the structural stability and electronic properties of lead-free halide double perovskite Cs$_2$KInI$_6$ using first-principles calculations and the machine-learned interatomic potential MACE.
Phonon calculations reveal that the high-symmetry cubic phase is dynamically unstable.
To find lower-energy dynamically stable configurations, we applied a genetic algorithm along unstable phonon eigenmodes, yielding 42 dynamically stable candidates according to finite-displacement phonon calculations with MACE.
A subset of structures was selected for further validation by DFPT to address possible MLIP inaccuracies. In two cases, structures predicted to be dynamically stable by the MLIP were found to remain dynamically unstable at the DFPT level, indicating that further DFPT validation is required for the remaining candidate structures. 
We focus on four dynamically stable phases: P$\bar{3}$, I$\bar{4}$2m, Cmc$2_1$, and P$\bar{1}$, classified by the coordination of In and K cations and their octahedral connectivity.
These phases are thermodynamically metastable based on their calculated energy above the convex hull. However, they lie within the threshold limit of 25 meV/atom for the iodides.
The most stable phases, I$\bar{4}$2m and Cmc$2_1$, are not double perovskites, as In is tetrahedrally coordinated and the K environment is ambiguous.
P$\bar{3}$ is a double perovskite, whereas P$\bar{1}$ is less clearly so: In is octahedrally coordinated but K is not.
Electronic-structure calculations show that distortions increase the band gap relative to the cubic phase and have changed it from direct to indirect in two cases.
The cubic phase has a direct band gap of 1.94~eV.
The double-perovskite P$\bar{3}$ phase has an indirect band gap of 2.29~eV.
The P$\bar{1}$ phase could be an interesting candidate among the stable phases for photovoltaics, retaining a direct band gap which is lower than the discussed stable phases.
Overall, Cs$_2$KInI$_6$ exhibits multiple candidate dynamically stable phases identified from phonon calculations using the machine-learned interatomic potential MACE, several of which are further confirmed by DFPT. These phases range from double to non-double-perovskites, with varying octahedral connectivity in the double-perovskite polymorphs.
These transformations significantly modify the electronic properties, clarifying structure–property relationships in halide double perovskites and demonstrating the effectiveness of combining machine-learned interatomic potentials with genetic algorithms to discover novel stable phases in complex materials.

\begin{acknowledgements}
G.-M.R. is Research Director of the Fonds de la Recherche Scientifique - FNRS.
S. P. is a Research Associate of the Fonds de la Recherche Scientifique - FNRS.
This work was supported by the Fonds de la Recherche Scientifique - FNRS under Grants number T.0183.23 (PDR) and T.W011.23 (PDR-WEAVE). 
This publication was supported by the Walloon Region in the strategic axe FRFS-WEL-T.
V. T. acknowledges support from the FRS-FNRS through a FRIA Grant.
Computational resources have been provided by the Consortium des Équipements de Calcul Intensif (CÉCI), funded by the Fonds de la Recherche Scientifique de Belgique (F.R.S.-FNRS) under Grant No.~2.5020.11 and by the Walloon Region.
The present research benefited from computational resources made available on the {Tier-1} supercomputer of the F\'ed\'eration Wallonie-Bruxelles, infrastructure funded by the Walloon Region under grant agreement n\textsuperscript{o}1117545.
\end{acknowledgements}

\bibliography{Bibliography}

\end{document}

\title{SI: From Symmetry to Stability: Structural and Electronic Transformation in Cs$_2$KInI$_6$}
%\date{\today}
\author{Mohammad Bakhsh$^1$}
%\email{mohammad.bakhsh@uclouvain.be}
\author{Victor Trinquet$^1$}
\author{Rogério Almeida Gouvêa$^1$}
\author{Gian-Marco Rignanese$^{1,2}$}
\author{Samuel Ponc\'e$^{1,2}$}
%\email{samuel.ponce@uclouvain.be}

\affiliation{$^1$UCLouvain, Institute of Condensed Matter and Nanosciences (IMCN), Chemin des \'Etoiles~8, B-1348 Louvain-la-Neuve, Belgium}
\affiliation{$^2$WEL Research Institute, avenue Pasteur 6, 1300 Wavre, Belgium.}

\pacs{}

\maketitle

\renewcommand{\figurename}{Figure}
\renewcommand{\thefigure}{S\arabic{figure}}
\renewcommand{\theequation}{S\arabic{equation}}
\renewcommand{\tablename}{Table}
\renewcommand{\thetable}{S\arabic{table}}

%\FloatBarrier
\begin{table*}[th]
    \caption{\label{tab:1sup}Polymorphs of Cs$_2$KInI$_6$ which are stable according to MACE, with the energy above the convex hull ($E_{hull}$ in meV/atom), band gap ($E_g$ in eV), and stability predicted with DFPT.}
    \begin{ruledtabular}
    %\resizebox{\textwidth}{!}{
    \begin{scriptsize}
    \begin{tabular}{c l r r r r r }
        id & \multicolumn{2}{c}{Space group} & $N_\mathrm{at}$ & $E_{hull}$ & $E_g$ & Stability with DFPT\\
    \hline

    00 & Fm$\bar{3}$m  & 225 &               10 & 55 &           1.24 & unstable  \\
    01 &           P2 & 3 &               10 &    46 &             1.76  & stable  \\  
    02 &     P$\bar{1}$ & 2 &               10 &  33 &           1.95 & stable  \\
    03 & P$\bar{3}$1m & 162 &               10 &  33 &           1.95 & unstable \\ 
    04 &        C2/m & 12 &               10 &    30 &           1.60 & stable \\ 
    05 &   P$\bar{3}$ & 147 &               10 &  24 &          1.66 & stable \\
    06 &           C2 & 5 &               10 &    21 &            2.24    & stable \\
    07 & I$\bar{4}$2m & 121 &               10 &  21 &           2.24  & stable \\  
    08 &     P$\bar{1}$ & 2 &               20 &  48 &           1.26    & -\\
    09 &    P2$_1$/c    &14 &              20 &   41 &           1.58    & unstable \\
    10 &     P$\bar{1}$ & 2 &               20 &  41 &           1.19 & -\\ 
    11 &     P$\bar{1}$ & 2 &               20 &  38 &           1.78 & -\\
    12 &     P$\bar{1}$ & 2 &               20 &  38 &           2.00 & -\\
    13 &     P$\bar{1}$ & 2 &               20 &  37 &           1.82 & -\\
    14 &     P$\bar{1}$ & 2 &               20 &   37 &           1.93 & -\\
    15 &     P$\bar{1}$ & 2 &               20 &   37 &           1.78 & -\\
    16 &     P$\bar{1}$ & 2 &               20 &   36 &           1.98 & -\\
    17 &     P$\bar{1}$ & 2 &               20 &   35 &           1.85 & -\\
    18 &     P$\bar{1}$ & 2 &               20 &   35 &           1.83 & -\\
    19 &     P$\bar{1}$ & 2 &               20 &   33 &           1.83 & -\\
    20 &     P$\bar{1}$ & 2 &               20 &   31 &           2.16 & -\\
    21 &     P$\bar{1}$ & 2 &               20 &   31 &           2.00 & -\\
    22 &     P$\bar{1}$ & 2 &               20 &   30 &           2.08 & -\\
    23 &     P$\bar{1}$ & 2 &               20 &   30 &           1.94 & -\\
    24 &     P$\bar{1}$ & 2 &               20 &   27 &           2.08 & -\\
    25 &     P$\bar{1}$ & 2 &               20 &   27 &           1.87 & -\\
    26 &     P$\bar{1}$ & 2 &               20 &   26 &           1.97 & -\\
    27 &     P$\bar{1}$ & 2 &               20 &   25 &           2.37 & -\\
    28 &     P$\bar{1}$ & 2 &               20 &   24 &           2.08 & -\\
    29 &     P$\bar{1}$ & 2 &               20 &   24 &           1.67 & -\\
    30 &     P$\bar{1}$ & 2 &               20 &   24 &           2.20 & -\\
    31 &     P$\bar{1}$ & 2 &               20 &   23 &           1.68 & -\\
    32 &        Cmcm & 63 &               20 &     21 &           2.15    & -\\
    33 &     P$\bar{1}$ & 2 &               20 &   21 &           1.89  & -\\
    34 &        C2/m & 12 &               20 &     19 &           2.31     & -\\
    35 &     P$\bar{1}$ & 2 &               20 &   19 &           2.31   & -\\
    36 &        Ama2 & 40 &               20 &     15 &           2.28  & -\\
    37 &    Cmc2$_1$ & 36 &               20 &     13 &           2.26  & stable \\ 
    38 &     P$\bar{1}$ & 2 &               40 &   31 &           1.93  & -\\
    39 &     P$\bar{1}$ & 2 &               40 &   29 &           2.25  & -\\
    40 &     P$\bar{1}$ & 2 &               40 &   27 &           1.36  & -\\
    41 &           Pc & 7 &               40 &     15 &           2.28    & stable\\
    42 &     P$\bar{1}$ & 2 &               80 &   25 &           1.35  & stable \\ 
    \end{tabular}
    \end{scriptsize}
    \end{ruledtabular}
\end{table*}
\FloatBarrier

% In the preamble:
% \usepackage{siunitx}
% (optional but helpful) \sisetup{detect-weight=true, detect-family=true}

\begin{table*}[tb]
    \caption{\label{tab:2sup}Polymorphs of Cs$_2$KInI$_6$ unstable according to MACE and DFPT, with energy above the convex hull ($E_{hull}$ in meV/atom), and band gap ($E_g$ in eV).}
    \begin{ruledtabular}
    \begin{tabular}{c  l r  r  r  r }
        id & \multicolumn{2}{c}{Space group} & $N_\mathrm{at}$ & $E_{hull}$ & $E_g$\\
        \hline
        00 & Fm$\bar{3}$m & 225 & 10 & 55 & 1.24 \\
        43 & R$\bar{3}$ &148 & 10 & 38 & 1.53 \\
        44 & C2/m & 12 & 10 & 33 & 1.95   \\
        45 & C2/m & 12 & 10 & 26 & 1.48 \\
        46 & Fmm2 & 42 & 10 & 21 & 2.13  \\
        47 & Cm & 8 & 10 & 21 & 2.14 \\
    \end{tabular}
    \end{ruledtabular}
\end{table*}

\begin{figure*}[!htbp]
    \centering
    \includegraphics[width=0.8\linewidth]{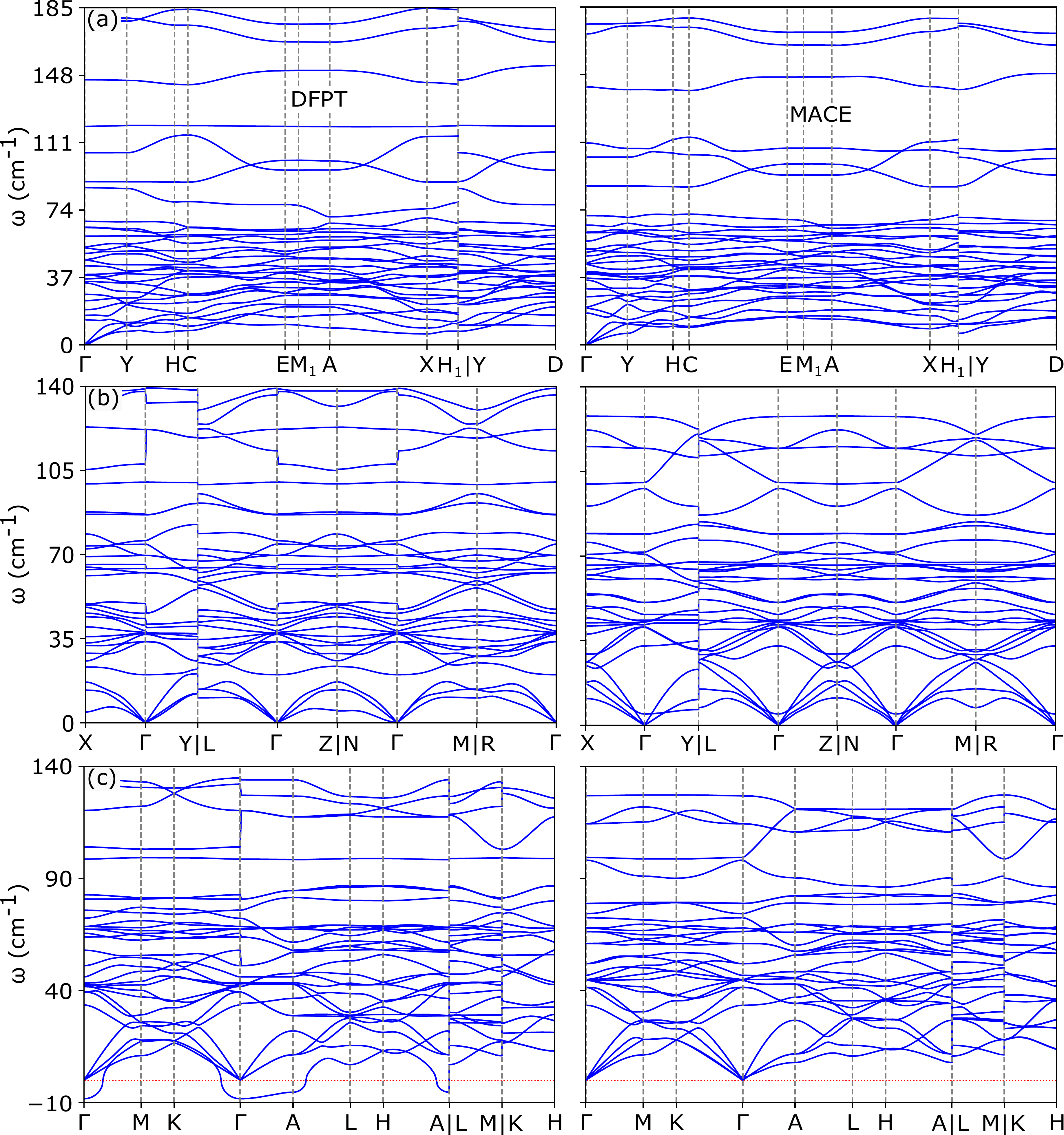}
     \caption{\label{fig:1sup}
     Phonon dispersions from DFPT (left) and finite displacement with MACE (right) for \ch{Cs2KInI6} phases with id numbers from Table~\ref{tab:1sup}. 
      All MACE calculations use a displacement of $0.01\,\text{\AA}$. 
     (a) id=01 with $4 \times 8 \times 4$ \textbf{k}-grid and $2 \times 4 \times 2$ \textbf{q}-grid (DFPT), and a $4 \times 7 \times 3$ supercell (MACE).
     (b) id=02 with $8 \times 6 \times 8$ \textbf{k}-grid and $4 \times 3 \times 4$ \textbf{q}-grid (DFPT), and $5 \times 4 \times 5$ supercell (MACE).
     (c) id=03 with $4 \times 4 \times 4$ \textbf{k}-grid and $2 \times 2 \times 2$ \textbf{q}-grid (DFPT), and a $4 \times 4 \times 4$ supercell (MACE).}
\end{figure*}

\begin{figure*}[ht]
    \centering
    \includegraphics[width=0.9\linewidth]{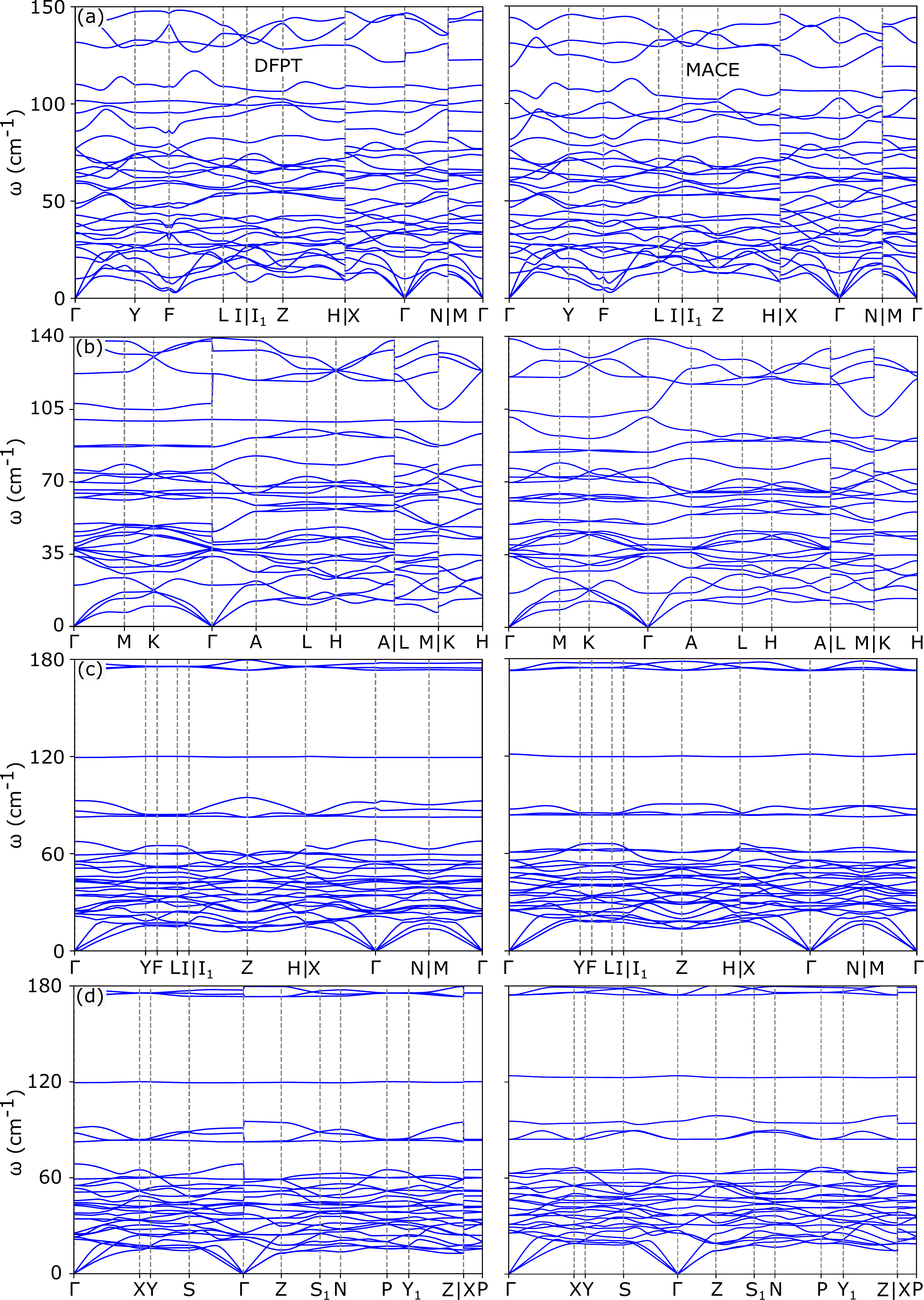}
     \caption{\label{fig:2sup} 
     Phonon dispersions from DFPT (left) and finite displacement with MACE (right) for \ch{Cs2KInI6} phases with id numbers from Table~\ref{tab:1sup}. 
      All MACE calculations use a displacement of $0.01\,\text{\AA}$. 
     (a) id=04 with $4 \times 4 \times 4$ \textbf{k}-grid and $2 \times 2 \times 2$ \textbf{q}-grid (DFPT), and a $5 \times 5 \times 5$ supercell (MACE).
     (b) id=05 with $4 \times 4 \times 4$ \textbf{k}-grid and $2 \times 2 \times 2$ \textbf{q}-grid, and a $5 \times 5 \times 5$ supercell (MACE).
     (c) id=06 with $4 \times 4 \times 6$ \textbf{k}-grid and $2 \times 2 \times 3$ \textbf{q}-grid (DFPT), and a $3 \times 3 \times 5$ supercell (MACE).
     (d) id=07 with $3 \times 3 \times 3$ \textbf{k}-grid and $2 \times 2 \times 2$ \textbf{q}-grid (DFPT), and a $5 \times 5 \times 5$ supercell (MACE).}
\end{figure*}

\begin{figure*}[!htbp]
    \centering
    \includegraphics[width=0.9\linewidth]{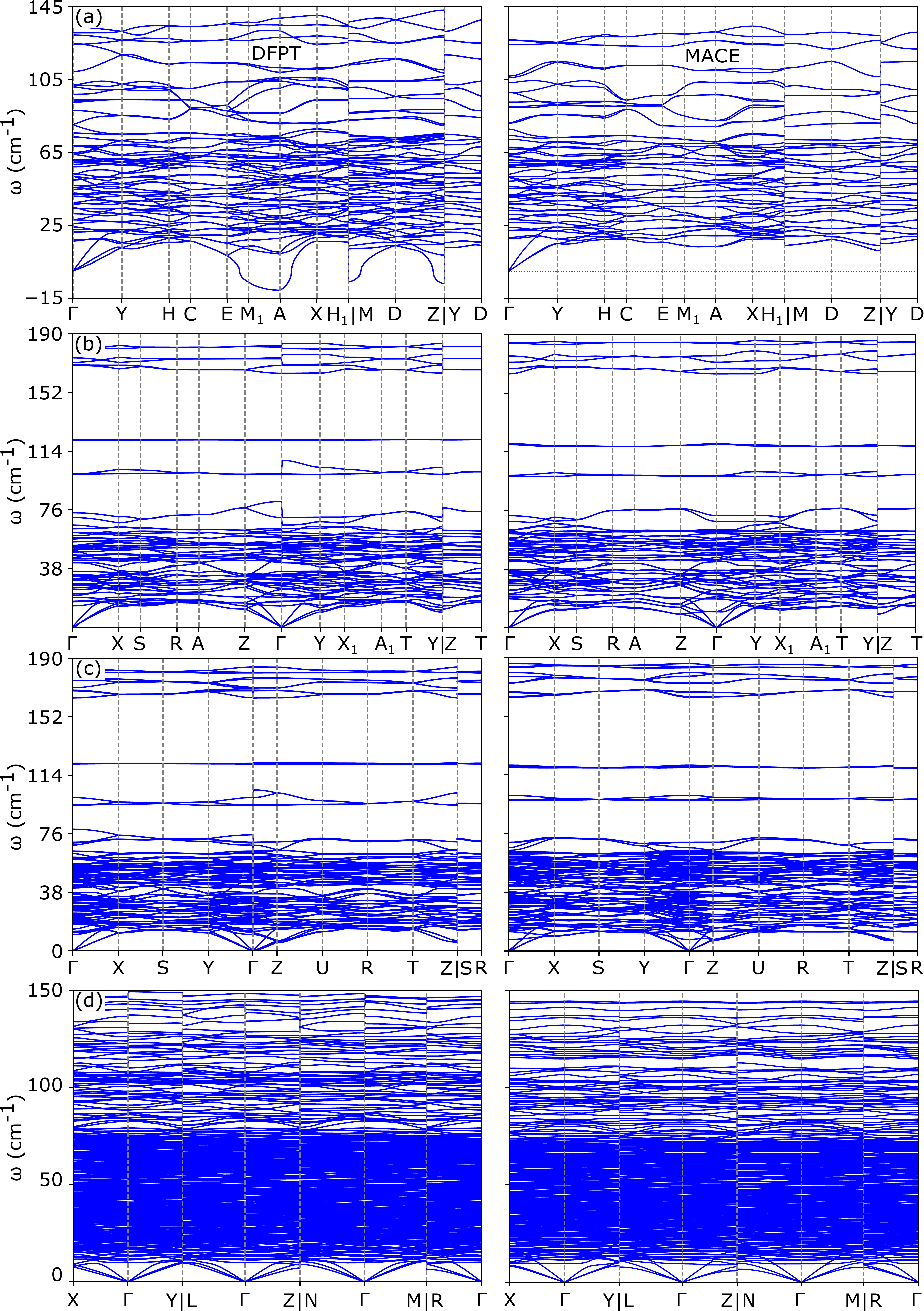}
     \caption{Phonon dispersions from DFPT (left) and finite displacement with MACE (right) for \ch{Cs2KInI6} phases with id numbers from Table~\ref{tab:1sup}. 
      All MACE calculations use a displacement of $0.01\,\text{\AA}$. 
     (a) id=09 with $4 \times 2 \times 4$ \textbf{k}-grid and $2 \times 1 \times 2$ \textbf{q}-grid (DFPT), and a $4 \times 3 \times 4$ supercell (MACE).
     (b) id=37 with $3 \times 3 \times 3$ \textbf{k}-grid and $2 \times 2 \times 2$ \textbf{q}-grid (DFPT), and a $3 \times 3 \times 3$ supercell (MACE).
     (c) id=41 with $4 \times 4 \times 2$ \textbf{k}-grid and $2 \times 2 \times 2$ \textbf{q}-grid (DFPT), and a $4 \times 4 \times 3$ supercell (MACE).
     (d) id=42 with $2 \times 2 \times 2$ \textbf{k}-grid and $1 \times 1 \times 1$ \textbf{q}-grid (DFPT), and a $3 \times 3 \times 3$ supercell (MACE).}
    \label{fig:3sup}
\end{figure*}

\begin{figure*}[!htbp]
    \centering
    \includegraphics[width=\linewidth]{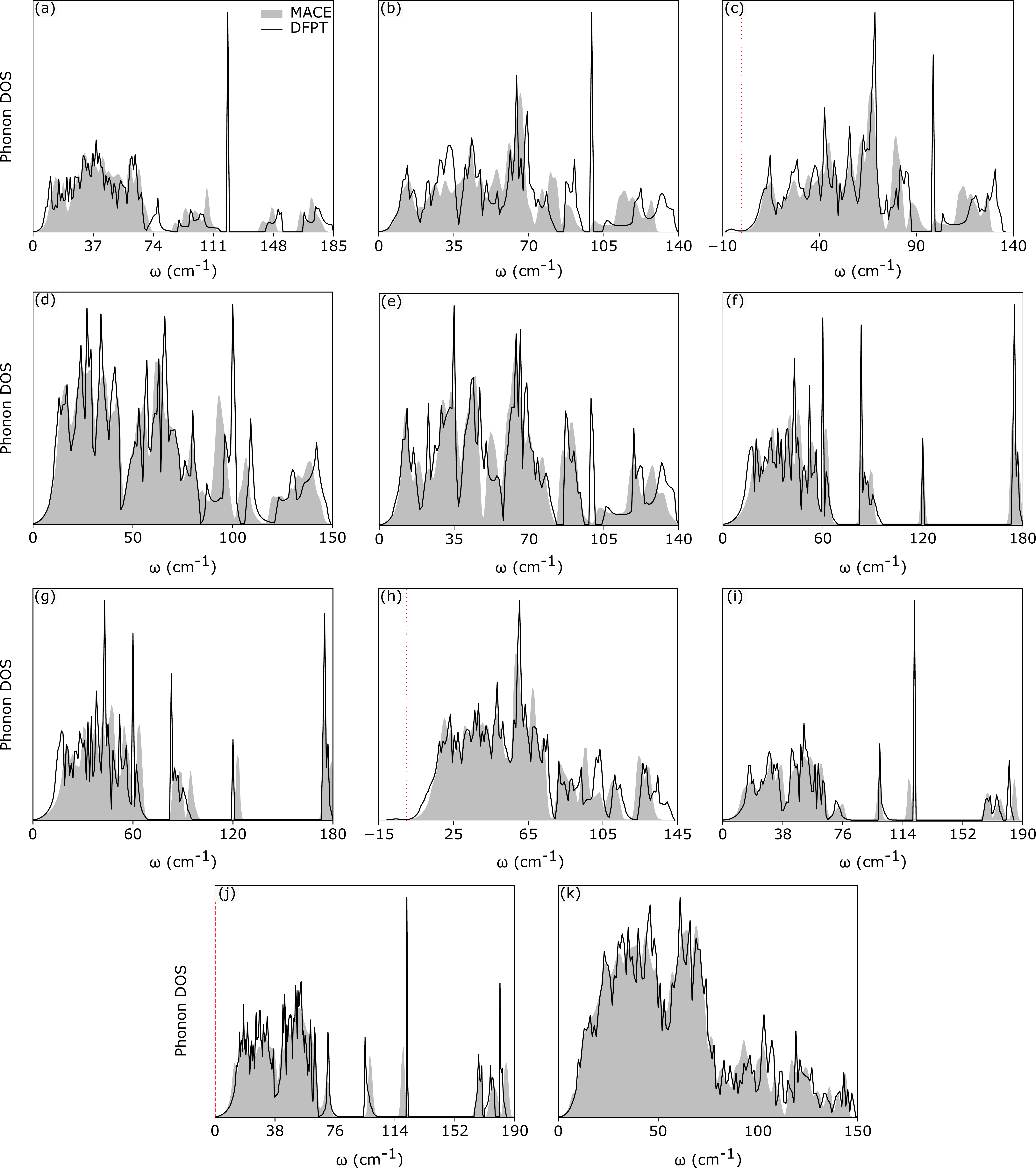}
     \caption{Phonon density of states (PDOS) obtained from DFPT and the finite-displacement method using MACE for \ch{Cs2KInI6} phases with id numbers listed in Table~\ref{tab:1sup}. The corresponding phonon band dispersions are shown in Figures~\ref{fig:1sup}, \ref{fig:2sup}, and \ref{fig:3sup}. A $40 \times 40 \times 40$ q-point grid was used to compute the PDOS, except for the structures with 40 (id=41) and 80 (id=42) atoms, for which $30 \times 30 \times 30$ and $15 \times 15 \times 15$ grids were used, respectively. The PDOS were normalized by the area under the curve within a common frequency range shared by the DFPT and MACE results.
     (a) id=01, (b) id=02, (c) id=03, (d) id=04, (e) id=05, (f) id=06, (g) id=07, (h) id=09, (i) id=37, (j) id=41, (k) id=42.}
    \label{fig:4sup}
\end{figure*}

\begin{figure*}[!htbp]
    \centering
    \includegraphics[width=0.9\linewidth]{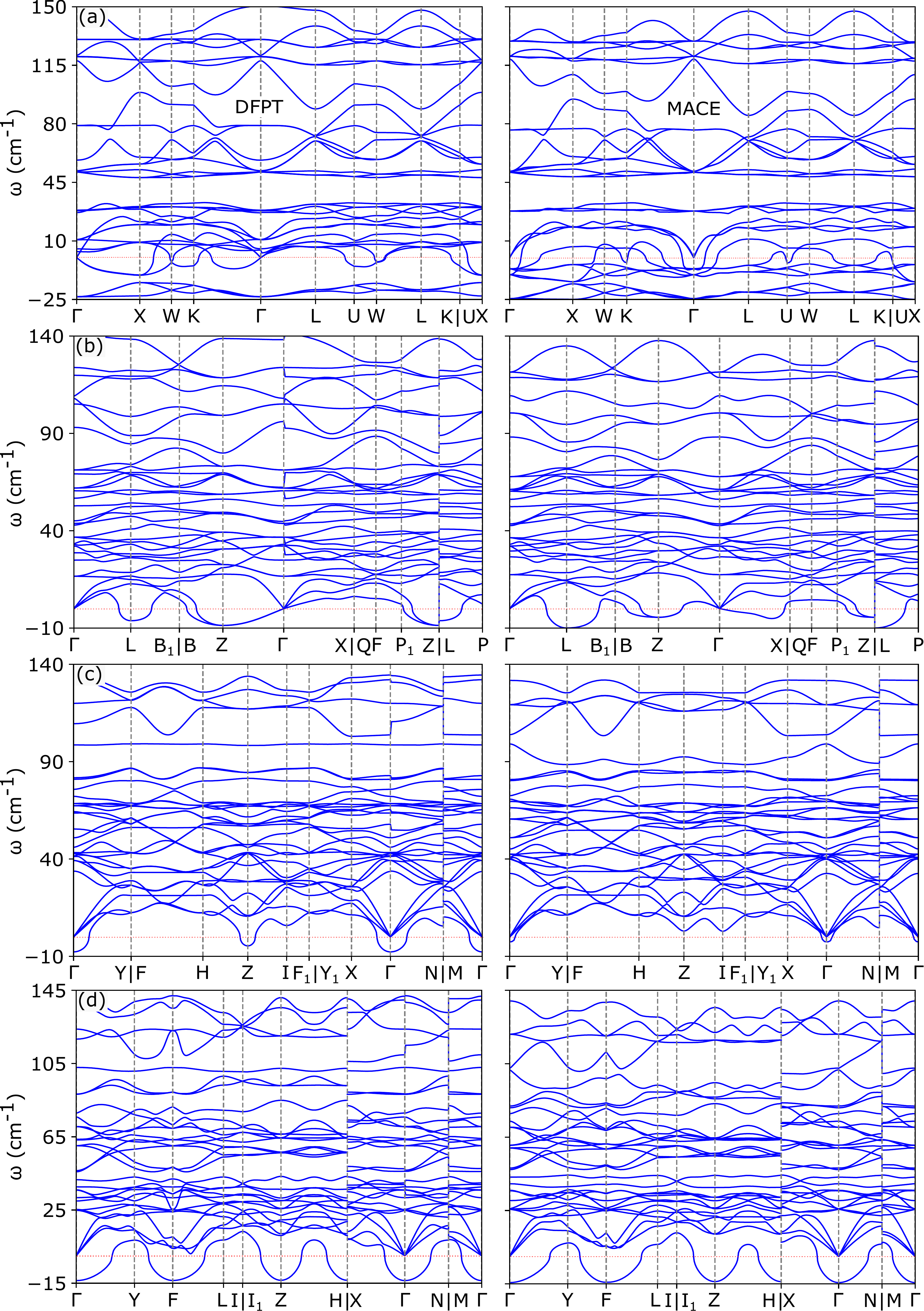}
     \caption{Phonon dispersions from DFPT (left) and finite displacement with MACE (right) for \ch{Cs2KInI6} phases with id numbers from Table~\ref{tab:2sup}. 
      All MACE calculations use a displacement of $0.01\,\text{\AA}$. 
     (a) id=00 with $3 \times 3 \times 3$ \textbf{k}-grid and $2 \times 2 \times 2$ \textbf{q}-grid (DFPT), and a $3 \times 3 \times 3$ supercell (MACE).
     (b) id=43 with $3 \times 3 \times 3$ \textbf{k}-grid and $2 \times 2 \times 2$ \textbf{q}-grid (DFPT), and a $5 \times 5 \times 5$ supercell (MACE).
     (c) id=44 with $6 \times 6 \times 4$ \textbf{k}-grid and $3 \times 3 \times 2$ \textbf{q}-grid (DFPT), and $6 \times 6 \times 4$ supercell (MACE).
     (d) id=45 with $4 \times 4 \times 4$ \textbf{k}-grid and $2 \times 2 \times 2$ \textbf{q}-grid (DFPT), and a $5 \times 5 \times 5$ supercell (MACE).}
    \label{fig:5sup}
\end{figure*}

\begin{figure*}[!htbp]
    \centering
    \includegraphics[width=0.8\linewidth]{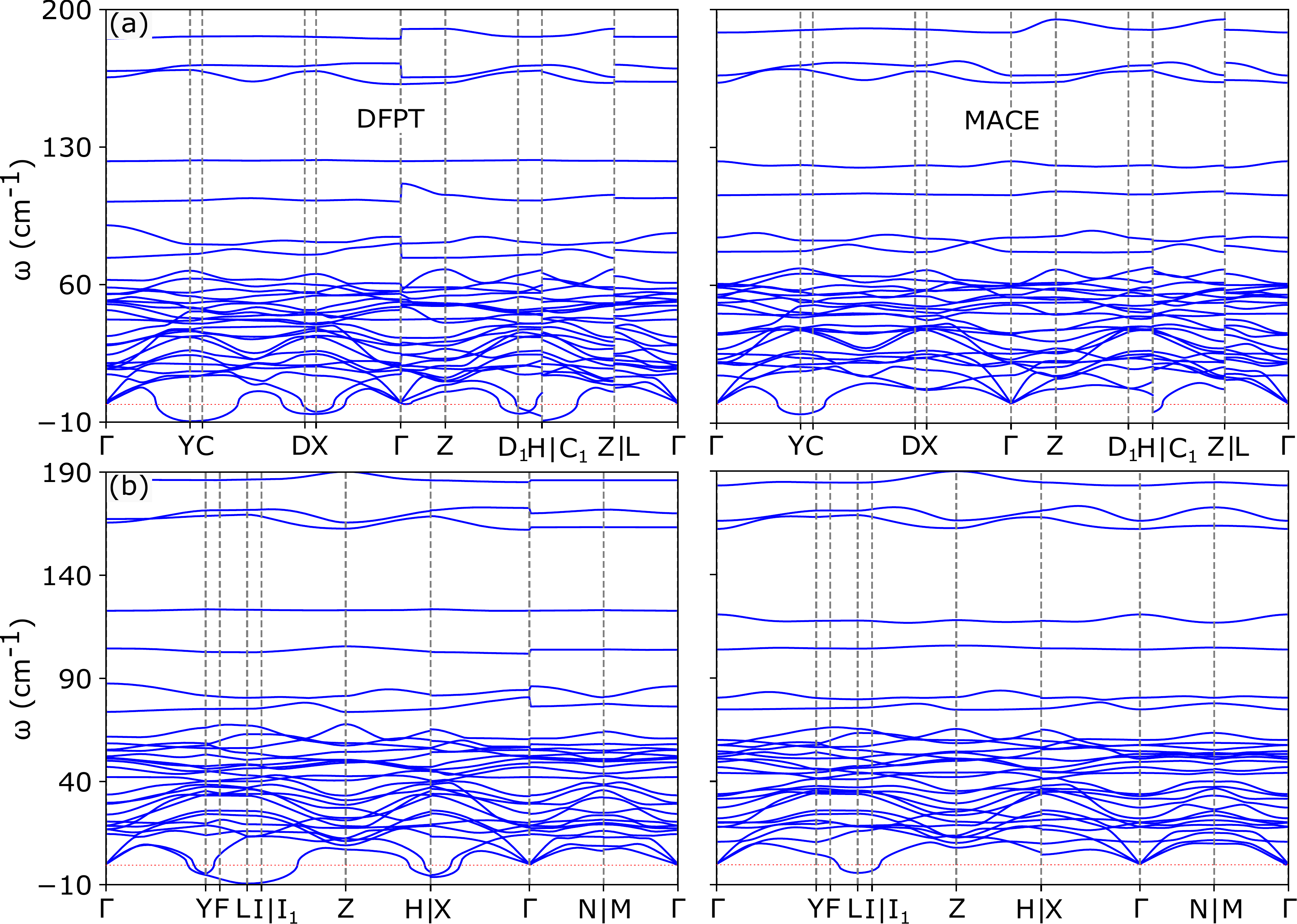}
     \caption{Phonon dispersions from DFPT (left) and finite displacement with MACE (right) for \ch{Cs2KInI6} phases with id numbers from Table~\ref{tab:2sup}. 
      All MACE calculations use a displacement of $0.01\,\text{\AA}$. 
     (a) id=46 with $4 \times 4 \times 6$ \textbf{k}-grid and $2 \times 2 \times 3$ \textbf{q}-grid (DFPT), and a $5 \times 5 \times 7$ supercell (MACE).
     (b) id=47 with $6 \times 6 \times 4$ \textbf{k}-grid and $3 \times 3 \times 2$ \textbf{q}-grid (DFPT), and a $5 \times 5 \times 3$ supercell (MACE).}
    \label{fig:6sup}
\end{figure*}

\begin{figure*}[!htbp]
    \centering
    \includegraphics[width=0.8\linewidth]{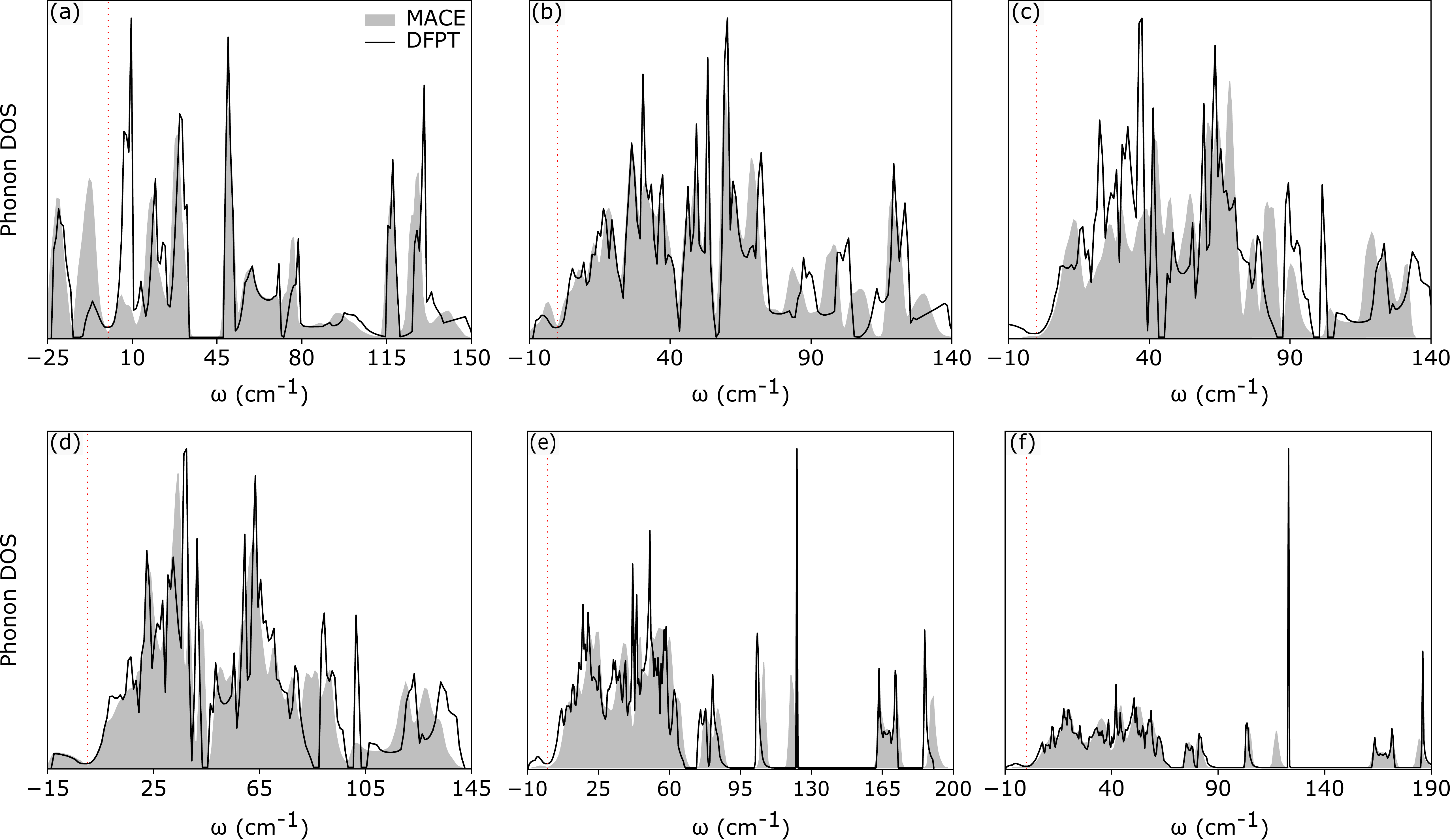}
     \caption{Phonon density of states (PDOS) obtained from DFPT and the finite-displacement method using MACE for \ch{Cs2KInI6} phases with id numbers listed in Table~\ref{tab:2sup}. The corresponding phonon band dispersions are shown in Figures~\ref{fig:5sup} and \ref{fig:6sup}. A $40 \times 40 \times 40$ q-point grid was used to compute the PDOS. The PDOS were normalized by the area under the curve within a common frequency range shared by the DFPT and MACE results.
     (a) id=00, (b) id=43, (c) id=44, (d) id=45, (e) id=46, (f) id=47.}
    \label{fig:7sup}
\end{figure*}

\begin{figure}[!htbp] 
    \centering
    \includegraphics[width=0.8\linewidth]{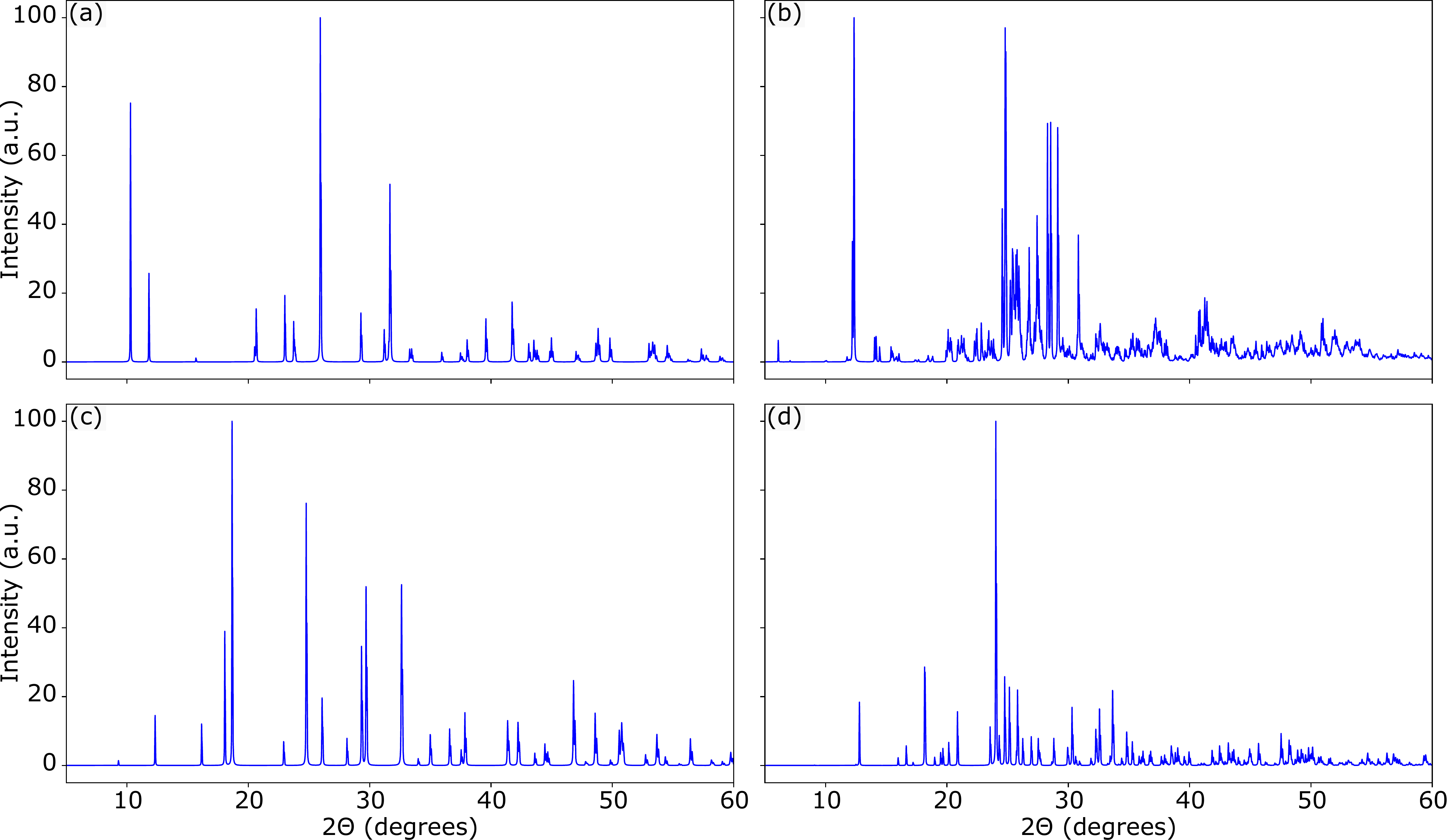}
     \caption{X-ray powder diffractogram of four dynamically stable \ch{Cs2KInI6} phases, obtained from VESTA with X-rays of wave length 1.54~\AA.
      (a) P$\bar{3}$ (147),(b) P$\bar{1}$ (2), (c) I$\bar{4}$2m (121), (d) Cmc$2_1$ (36).}
    \label{fig:8sup}
\end{figure} 

\begin{figure}[!htbp] 
    \centering
    \includegraphics[width=\linewidth]{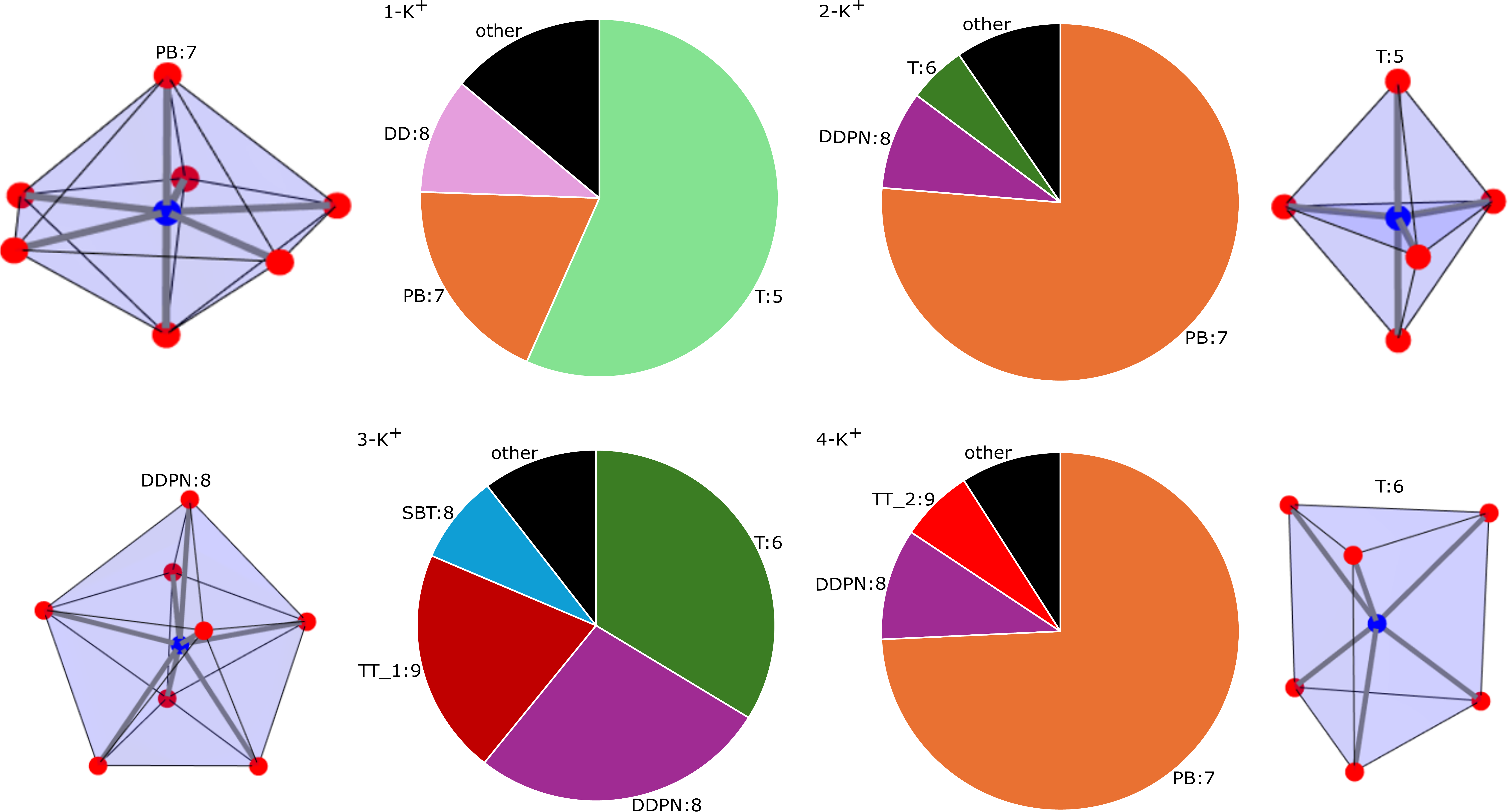}
    \caption{\label{fig:9sup} 
     Analysis of the coordination environments of K cations for P$\bar{1}$ (2): T:5 = Trigonal bipyramid, PB:7 = Pentagonal bipyramid, T:6 = Trigonal prism, DD:8 = Dodecahedron with triangular faces, DDPN:8 = Dodecahedron with triangular faces – p2345 plane normalized, TT\_1:9 = Tricapped triangular prism (three square-face caps), TT\_2:9 = Tricapped triangular prism (two square-face caps and one triangular-face cap), SBT:8 = Square-face bicapped trigonal prism.}
\end{figure}

\begin{table*}[!htbp]
    \caption{\label{tab:3sup}Lattice parameters ($a$, $b$, and $c$ in \AA; angles $\alpha$, $\beta$, and $\gamma$ in $^\circ$) of the dynamically stable polymorphs of \ch{Cs2KInI6}, computed using PBEsol exchange-correlation functional.}
    \begin{ruledtabular}
    \begin{tabular}{l r c r r r r r r}
         \multicolumn{2}{c}{Space group} & id & \multicolumn{1}{c}{$a$} & \multicolumn{1}{c}{$b$} & \multicolumn{1}{c}{$c$} & \multicolumn{1}{c}{$\alpha$} & \multicolumn{1}{c}{$\beta$} & \multicolumn{1}{c}{$\gamma$} \\
    \hline
    \rule{0pt}{3ex}Fm$\bar{3}$m & 225 &00 & 12.332  & 12.332  & 12.332  & 90.0  & 90.0  & 90.0 \\
    P$\bar{3}$ & 147 &05 & 8.267 & 8.267 & 8.381 & 90.0 & 90.0 & 120.0 \\
    I$\bar{4}$2m & 121 &07 & 7.575 & 7.575 & 18.206 & 90.0 & 90.0 & 90.0 \\
    Cmc$2_1$ & 36 &37& 10.280 & 18.919 & 9.881 & 90.0 & 90.0 & 90.0 \\
    P$\bar{1}$ & 2 &42 & 16.767 & 16.826 & 17.221 & 91.9 & 118.5 & 118.9 \\       
    \end{tabular}
    \end{ruledtabular}
\end{table*}

\begin{figure}[!htbp]
    \centering
    \includegraphics[width=0.9\linewidth]{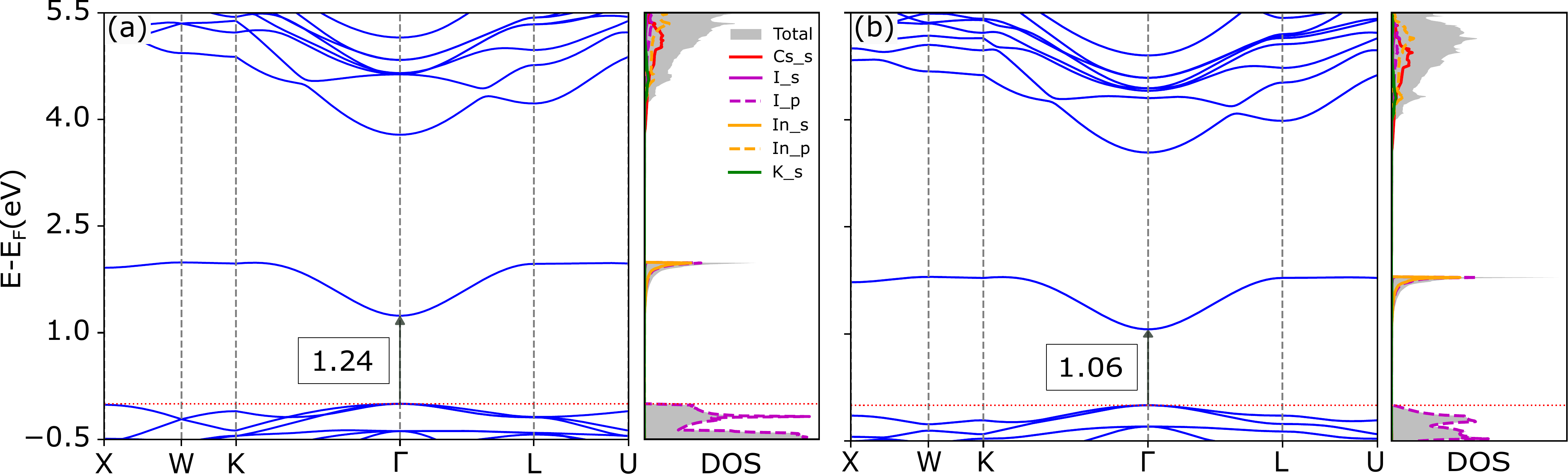}
    \caption{\label{fig:10sup}Electronic band structures and densities of states of \ch{Cs2KInI6} for the high-symmetry unstable Fm$\bar{3}$m (225) phase calculated using (a) PBE and (b) PBE+SOC.
    The energies are expressed with respect to the Fermi level ($E_{\rm F}$), located at the valence band maximum.}
\end{figure}

\begin{figure*}[!htbp]
    \centering
    \includegraphics[width=0.9\linewidth]{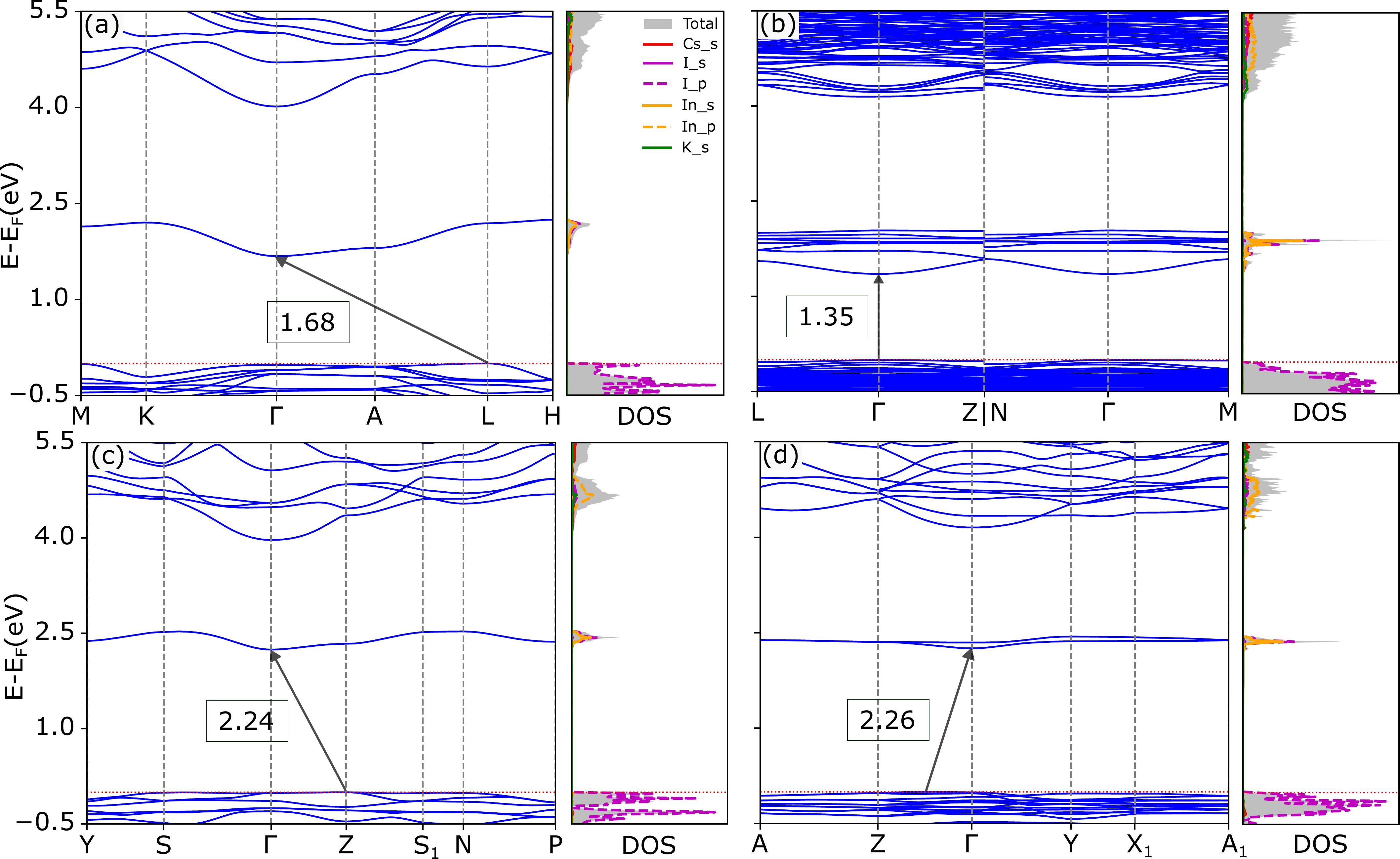}
    \caption{\label{fig:11sup} Electronic band structures and densities of states of dynamically stable phases of \ch{Cs2KInI6} calculated at the PBE level: (a) P$\bar{3}$, (b) P$\bar{1}$, (c) I$\bar{4}$2m, and (d) Cmc$2_1$.
    The energies are expressed with respect to the Fermi level ($E_{\rm F}$), located at the valence band maximum.}
\end{figure*}

\begin{figure*}[!htbp]
    \centering
    \includegraphics[width=\linewidth]{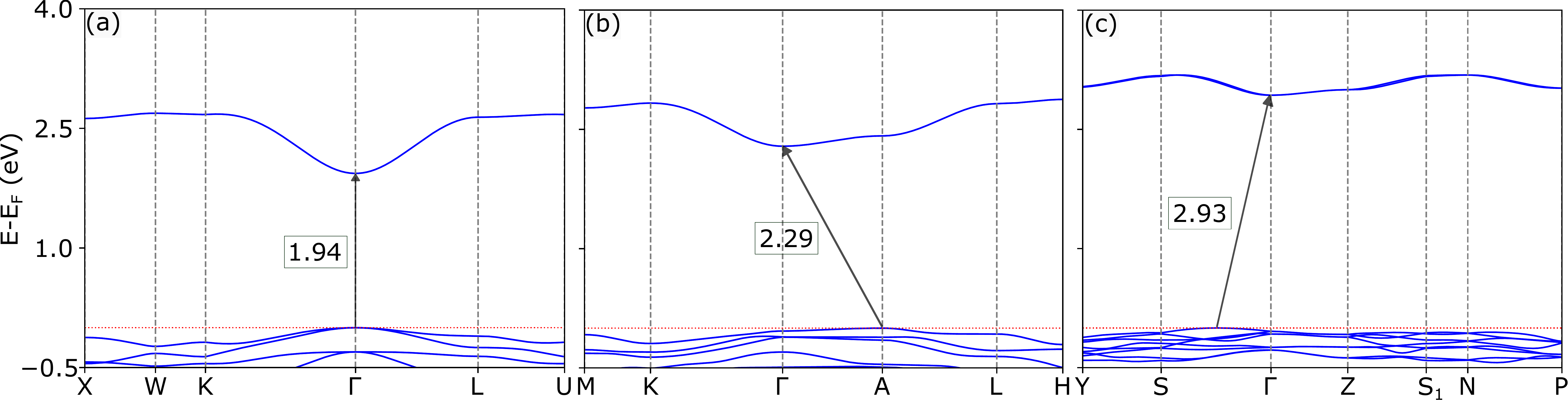}
    \caption{\label{fig:11sup}Electronic band structures of \ch{Cs2KInI6} phases calculated using HSE06+SOC and Wannier interpolation: (a)Fm$\bar{3}$m, (b) P$\bar{3}$, and (c) I$\bar{4}$2m.
    The energies are expressed with respect to the Fermi level ($E_{\rm F}$), located at the valence band maximum.}
\end{figure*}

%
%\bibliographystyle{apsrev4-1}
%\bibliography{Bibliography}